\documentclass[final,5p,times,twocolumn,sort&compress]{elsarticle}

\usepackage{amsmath,amssymb,graphicx,bm}
\usepackage{enumitem}
\usepackage{cuted}
\usepackage[colorlinks=true,linkcolor=blue,citecolor=blue,urlcolor=blue]{hyperref}
\usepackage{cleveref}

\usepackage{xcolor}

\makeatletter
\def\ps@pprintTitle{%
 \let\@oddhead\@empty 
 \let\@evenhead\@empty
 \def\@oddfoot{}%
 \let\@evenfoot\@oddfoot}
 \makeatother

\newcommand{\x}{\mathbf{x}}
\newcommand{\f}{\mathbf{f}}
\newcommand{\rv}{\mathbf{r}}
\newcommand{\lv}{\mathbf{l}}

\newcommand{\J}{\mathbf{J}}
\newcommand{\A}{\mathbf{A}}

\newcommand{\PP}{\mathbf{P}}

\newcommand{\xii}{\boldsymbol{\xi}}

\newcommand{\nabl}{\boldsymbol{\nabla}}

\newcommand{\mw}[1]{\langle #1 \rangle}

\begin{document}

   \title{Life as Non-Normal Chemical Accelerator}

    \author{Didier Sornette}
    \author{Virgile Troude}
    \address{Institute of Risk Analysis, Prediction and Management (Risks-X),
    Academy for Advanced Interdisciplinary Sciences,
    Southern University of Science and Technology, Shenzhen, China}

    \date{\today}

    \begin{abstract}
      Life is commonly described as a self-organized, far-from-equilibrium process that maintains internal order by consuming free energy and exporting entropy. This thermodynamic view underlies diverse theoretical frameworks---from autopoiesis and relational biology to autocatalytic sets and hypercycles---yet dissipation is typically treated as a necessary consequence of living organization rather than as a property shaped by its internal dynamics.
Here, through explicit calculations of biotic chemical reactions and empirical documentation, we show that living systems universally function as non-normal chemical accelerators. Their elevated entropy production emerges from the asymmetric and hierarchical architecture of their biochemical networks.
We introduce a general conceptual and mathematical framework in which biological structuration is understood as a dynamical property.
Characterized by asymmetric couplings and transient amplification despite asymptotic stability,  non-normal dynamics are shown to naturally generate kinetic acceleration, enhanced energy throughput, and phase-transition-like reorganizations without classical bifurcations. In this view, biological organization is not merely compatible with dissipation but actively structured to amplify free-energy flux and entropy export.
We support this perspective with empirical and theoretical evidence that biochemical networks generically give rise to intrinsically non-normal operators through non-reciprocal interactions and hierarchical design. This framework yields testable predictions for dissipation rates, robustness, and evolutionary design principles, and suggests a kinetic principle of evolution in which living systems preferentially construct increasingly non-normal reaction architectures, driving sustained amplification of chemical fluxes and entropy flow.
\vskip 0.3cm

  {\bf One sentence summary}: We show that life's defining thermodynamic signature---high entropy production---emerges not merely as a consequence of being far from equilibrium, but from a universal dynamical principle: the evolution of intrinsically non-normal biochemical architectures that actively amplify chemical fluxes.
    \end{abstract}

    \maketitle

The major theoretical frameworks developed to capture the essence of life---autopoiesis \cite{MaturanaVarela1980,VarelaMaturanaUribe1974}, 
Rosen's (M,R)-systems \cite{Rosen1958,Rosen1991}, Kauffman's autocatalytic sets \cite{Kauffman1986,HordijkSteel2004}, G\'anti's chemoton \cite{Ganti2003} and Eigen and Schuster's hypercycle \cite{Eigen1971,EigenSchuster1977,EigenSchuster1978}---form the historical core of theoretical biology, each emphasizing complementary aspects of organization, metabolism, information, and causation. Although these approaches appear at first to differ markedly, they share a common lineage tracing back to Schr\"odinger's What is Life?~\cite{Schrodinger_WhatIsLife}, which framed living organisms as systems that persist by importing free energy and exporting entropy, thereby sustaining organized, far-from-equilibrium states. 
Despite their differing emphases on metabolism, information, boundary formation, or relational closure, each of these frameworks captures a distinct facet of the same underlying phenomenon: the emergence of autonomous, self-producing processes that exploit environmental thermodynamic disequilibria, such as chemical and redox free-energy differences, to perform work, maintain organized structure, and enable growth, reproduction, and open-ended complexity over time. These foundational theories are now commonly discussed alongside nonequilibrium thermodynamic descriptions of dissipative structures \cite{NicolisPrigogine1977}, metabolism-first origin-of-life scenarios \cite{Wachtershauser1988}, information-first and RNA-world hypotheses \cite{Gilbert1986}, and more recent process-based, autonomy-centered, and information-theoretic formulations of life \cite{MorenoMossio2015,Deacon2011,WalkerDavies2013,Friston2010}, which together extend and refine this unifying view.

Accordingly, maintaining persistent low-entropy organization in an open system requires sustained free-energy throughput and entropy export; as levels of structural organization, regulation, and functional integration increase, so too must the associated energy fluxes and entropy release to the environment \cite{SchneiderKay1994}. The further such a system is maintained from thermodynamic equilibrium, through sustained chemical, electrochemical, or informational gradients, the larger the required fluxes of matter and energy and the higher the associated entropy production. In the linear nonequilibrium regime, this relationship can be made explicit: entropy production is a positive-definite quadratic form of the thermodynamic forces driving transport and reactions \cite{Onsager1931a,Onsager1931b,Qian2006}, implying that stronger maintained gradients and more ordered steady states unavoidably entail larger fluxes and higher rates of entropy production that must be exported outward \cite{Prigogine1967,deGrootMazur1984,NicolisPrigogine1977}.

Moreover, the dominant entropic cost of biological self-organization does not primarily arise from the one-time assembly of ordered components, but from the ongoing irreversibility of the processes that sustain and replicate them. For living systems, and even for comparatively simple organisms such as aerobic bacteria, the energetic cost of molecular assembly is small compared to the continuous entropy generated by metabolic dissipation and by self-replication itself, which produces long-lived, highly ordered copies that do not spontaneously relax back to equilibrium \cite{England2013}. Taken together, these considerations indicate that increasing biological complexity is inseparable from increasing free-energy consumption and entropy export, revealing a fundamental coupling in which organized living structure (and the internal constraints, regulation, and inference it entails) can emerge and persist only through enhanced dissipation and an increased flux of entropy into the environment.

While it is well established that such biological organization requires sustained dissipation, this classical thermodynamic view remains incomplete.  Here we establish a deeper physical basis for biological dissipation: living systems actively structure their internal dynamics in ways that systematically enhance free-energy throughput and entropy flux to their surroundings. Biological organization is thus not merely compatible with high dissipation but is progressively shaped to amplify it, as increasing levels of regulation, integration, and hierarchical organization enable faster and more effective exploitation of environmental free-energy gradients.
To formalize this idea, we introduce a general conceptual and mathematical framework in which biological ``structuration'' is understood not as static architecture but as a dynamical property of the underlying processes. We demonstrate that the intrinsically non-normal chemical dynamics characteristic of biological networks---arising from asymmetric, hierarchical, and non-reciprocal interactions---lead to accelerated reaction kinetics and enhanced energy throughput, with entropy production increasing sharply as a function of non-normality, even when asymptotic stability, system size, and stored energy are held fixed. In this framework, living systems do not rely on dissipation merely to preserve structure; instead, they engineer asymmetric couplings among reactions that permit transient amplification of fluxes, rapid kinetic acceleration, and elevated entropy production.

Rather than being generic chemical reactions, these non-normal biochemical architectures emerge as the primary generative cause of biological organization, defining an accelerated physico-chemical manifold sustained by continuous energy flows. Our theory explains the amplification signatures of widely observed biological motifs, including enzyme cycles, futile cycles, push-pull loops, covalent-modification cascades, membrane transport asymmetries, and metabolic modules exhibiting burst-like responses. A biotic-abiotic ``horse race'' further illustrates how non-normal organization confers decisive kinetic advantages over ordinary chemistry. From this perspective, life dissipates energy at high rates because its internal dynamical skeleton channels free energy into structured, accelerated work; entropy production thus emerges as a diagnostic consequence rather than a defining axiom. 

The framework yields clear, testable predictions: non-normal reaction networks should (i) dissipate energy faster under fixed driving forces, (ii) combine robustness to noise with heightened responsiveness, and (iii) exhibit reduced entropy production and slower kinetics when asymmetries are artificially suppressed. Together, these predictions outline a path toward experimentally validating life as a non-normal accelerator of far-from-equilibrium chemical fluxes, reframing the origin and persistence of life as a kinetic transition in which asymmetry and feedback transform passive chemistry into active, self-sustaining organization.


\section*{Formulation of Life's Non-Normal Reaction Architecture}


Biological chemistry operates predominantly in \emph{open systems maintained far from thermodynamic equilibrium}. Let
$\boldsymbol{\rho}(t) = (\rho_1(t), \ldots, \rho_M(t))$
denote the concentrations of molecular species participating in a biochemical network. Their dynamics can be written in the general form
$\dot{\boldsymbol{\rho}} = \f(\boldsymbol{\rho})$.
where $\f$ encodes reaction kinetics, transport, regulation, and exchange with the environment. Living systems maintain \emph{nonequilibrium steady states} (NESS) $\boldsymbol{\rho}^\ast$, sustained by continuous fluxes of energy and matter.

Considering small fluctuations $\mathbf{x}(t) = \boldsymbol{\rho}(t) - \boldsymbol{\rho}^\ast$
around such a NESS, linearization yields
\begin{equation}
\dot{\mathbf{x}} = \mathbf{A}\,\mathbf{x},
\label{frhwbgrq}
\end{equation}
where $\mathbf{A} = \nabla \f(\boldsymbol{\rho}^\ast)$ is the Jacobian (kinetic operator) of the biochemical network. The transient dynamics, response to perturbations, noise amplification, and kinetic acceleration of the system are therefore governed by the structure of $\mathbf{A}$.

Crucially, biochemical systems in living organisms operate far from equilibrium, sustained by chemical driving that breaks detailed balance and supports persistent fluxes. Recent analyses of primitive metabolic cycles and autocatalytic reaction networks explicitly demonstrate that even minimal biochemical architectures break detailed balance and generate non-reciprocal couplings that sustain directed fluxes and kinetic amplification \cite{OuazanReboul2023,Kauffman1986,HordijkSteel2004}. This nonequilibrium character, when linearized around a steady state, is naturally encoded in a non-normal dynamical operator $\mathbf{A}$, defined by the condition 
\begin{equation}
\mathbf{A}\mathbf{A}^\dagger \neq \mathbf{A}^\dagger \mathbf{A}.
\end{equation}

Non-normality arises generically from the inherent asymmetry of biological couplings. The influence of one species on another, via catalysis, regulation, or transport, is rarely matched by an equal reverse influence. These unavoidable asymmetries, driven by enzyme specificity, compartmentalization, and regulatory hierarchies, produce directional fluxes and ensure that the Jacobians governing biochemical dynamics are intrinsically non-normal, even 
when all eigenvalues have negative real parts that ensure stabllity.
At the cellular level, biochemical organization is deeply hierarchical: enzymes act on substrates, pathways feed into modules, and modules feed into cellular functions. Each hierarchical level introduces asymmetry in coupling strengths, time scales, and feedback directions, naturally generating strongly asymmetric or upper-triangular Jacobians,, which are canonical examples of non-normal operators.

While broken detailed balance explains the thermodynamic origin of steady-state fluxes and entropy production, it does not account for the kinetic efficiency of transient dynamics. Non-normality is the key missing element: it is a global property of $\mathbf{A}$ that generalizes the breakdown of detailed balance for a single interaction and captures the collective impact of asymmetric couplings on transient amplification and flux response. Thus, life's dynamical efficiency is governed not just by thermodynamic driving, but by the non-normal geometry of its underlying network.

A standard characterization of non-normality is the \emph{non-normality defect}
$\Delta(\mathbf{A}) \;=\; \|\mathbf{A}\mathbf{A}^\dagger-\mathbf{A}^\dagger\mathbf{A}\|$,
which vanishes if and only if $\mathbf{A}$ is normal. A complementary and often more directly dynamical measure is the \emph{eigenvector condition number}. If $\mathbf{A}$ is diagonalizable, $\mathbf{A}=\PP\boldsymbol{\Lambda} \PP^{-1}$, 
where the columns of $\mathbf{V}$ are right eigenvectors and $\boldsymbol{\Lambda}$ is the diagonal matrix of eigenvalues. The condition number
\begin{equation}
\kappa(\PP) \;=\; \|\PP\|\,\|\PP^{-1}\|
\label{qrtnthnh}
\end{equation}
quantifies eigenvector non-orthogonality and the potential for transient amplification: $\kappa(\PP)=1$ for unitary $\PP$ (orthonormal eigenvectors), while $\kappa(\PP)\gg 1$ indicates strong non-normality and large pseudospectral sensitivity. Equivalently, one may use the \emph{Hermitian part} $\mathbf{H}=(\mathbf{A}+\mathbf{A}^\dagger)/2$, since the instantaneous growth of the norm satisfies
$\frac{d}{dt}\|\mathbf{x}\|^2 \;=\; 2\,\mathbf{x}^\dagger \mathbf{H}\,\mathbf{x}$,
so that transient growth is possible whenever $\mathbf{H}$ has positive directions even if all eigenvalues of $\mathbf{A}$ lie in the stable half-plane.

Non-normality implies non-orthogonal eigenvectors and permits \emph{transient amplification}, fluctuation growth, noise-induced switching, and abrupt flux reorganizations despite asymptotic stability. These phenomena are widely observed in cellular regulation and metabolism. 
Many canonical biological motifs, such as phosphorylation-dephosphorylation cycles, futile cycles, push-pull loops, covalent-modification cascades, transcriptional regulation, and metabolic branch points, exhibit hallmark features of non-normal dynamics. These features include rapid transient amplification, high gain without instability, robustness coupled to sensitivity, and increased entropy production under fixed driving \cite{GoldbeterKoshland1981,Alon2007,Qian2006}. Non-normality is therefore not an added feature of biological chemistry; it is its \emph{default dynamical condition}.

Beyond intracellular chemistry to organismal behavior, locomotion converts metabolic free energy into directed transport, transforming diffusion-limited encounters into advection--reaction kinetics. From bacterial chemotaxis and ciliary swimming to fungal foraging, planktonic motion, and ecological dispersal, biological transport processes systematically break reciprocity and detailed balance \cite{Berg1972,Berg1993,LaugaPowers2009,Heaton2012,Kiorboe2008}.
Mathematically, advection--diffusion--reaction systems are governed by \emph{non-self-adjoint transport operators}. Advection and chemotactic drift introduce non-normal terms that produce transient growth, front sharpening, and enhanced uptake even when the spectrum indicates linear stability. Chemotactic feedback further couples motion and chemistry, steepening gradients and sustaining probability currents \cite{Purcell1977,Dusenbery1997,Dusenbery1999}.
Thus, motility represents a macroscopic manifestation of the same non-normal principle that accelerates intracellular chemistry. In both cases, asymmetric couplings transform steady driving into amplified fluxes and increased entropy production.


\section*{Non-normal acceleration of entropy production and reaction rates}

Here we develop the central dynamical mechanism that links non-normality to accelerated reaction kinetics and enhanced entropy production. We show that the geometric properties of non-normal operators, most notably eigenvector non-orthogonality, directly increase dissipation in nonequilibrium steady states (NESS) and enable strong transient amplification. These effects act in concert to lower effective kinetic barriers, so that elevated entropy production feeds back into faster chemical transitions. Together, they establish non-normality as a unifying dynamical mechanism through which biological systems amplify fluctuations, increase entropy flow, and accelerate chemical processes.

\subsection*{Entropy production enhanced by eigenvector overlap}

Extending (\ref{frhwbgrq}), we can describe general stochastic reaction-diffusion dynamics linearized around a steady state as
\begin{equation}    \label{eq:lin_master}
    \dot{\x} = \A\, \x + \xii,
    \qquad \mw{\xii(t)\xii^\top(t')} = \mathbf{I}\delta(t-t'),
\end{equation}
with $\A$ the kinetic Jacobian and $\mathbf{I}$ the identity matrix.
In the normal case, eigenmodes relax independently.
In the non-normal regime, however,
modes interact through the biorthogonal eigenvector overlaps $O_{ij}=(\rv_i\cdot\rv_j)(\lv_i\cdot\lv_j)$,
where $\lv_i$ and $\rv_i$ are respectively left and right eigenvectors of the Jacobian.
The steady-state entropy production rate \cite{Fyodorov2025} takes the general form
\begin{equation}
    \Phi = \sum_{i,j} O_{ij} \Lambda_{ij}, \qquad
    \Lambda_{ij}=\frac{\lambda_i(\lambda_i-\bar\lambda_j)}{\lambda_i+\bar\lambda_j},
    \label{hh3umju3}
\end{equation}
where the $\lambda_i$'s denote the eigenvalues of $\A$.
$  \Phi$ quantifies the irreversible circulation of probability currents,
the magnitude of excess heat released during slow operations,
and the geometrical strength of mode coupling induced by non-normality.
In an open metabolic chain,
increasing the non-normal condition number $\kappa$ (\ref{qrtnthnh}) enhances both the eigenvector-overlap matrix $O_{ij}$ and the entropy flux $\Phi$,
thereby amplifying the magnitude and frequency of noise-driven transitions.

If $\A$ is normal, the overlap matrix satisfies $O_{ij}=\delta_{ij}$.
For stationary nonequilibrium steady states (NESS) with real eigenvalues $\lambda_i=\bar{\lambda}_i$,
one has $\Lambda_{ii}=0$ and hence $\Phi=0$.
In this case, the steady-state entropy production vanishes.
By contrast, a non-zero steady-state entropy production rate arises 
only for non-normal operators $\A$, corresponding to housekeeping entropy: the system is maintained in a fixed nonequilibrium ensemble without time-dependent external driving via entropy production \cite{OonoPaniconi1998,HatanoSasa2001}.
For cyclic NESS, the eigenvalues $\lambda_i$ are complex and $\Phi$ is always positive, reflecting the continuous entropy production required to sustain cyclic processes. More generally, expression~(\ref{hh3umju3}) shows that, at fixed eigenvalues, entropy production increases with eigenvector non-orthogonality
and thus with non-normality. A contribution to excess entropy would arise only under a time-dependent external protocol that forces the steady-state distribution to evolve.

\subsection*{Transient amplification and dynamical ``barrier lowering''}

Non-normal operators exhibit large transient responses despite having all eigenvalues in the stable half-plane. The propagator norm obeys
\(
|e^{At}| \sim G_{\max} \propto \kappa^{2},
\),
where $\kappa$ denotes the condition number (\ref{qrtnthnh}) of the matrix $\PP$ 
whose columns are the right eigenvectors of $\A$; it is given by the ratio of the largest to the smallest singular values of $\PP$.
Thus, fluctuations are effectively amplified by a geometric factor controlled by non-normality.
As shown in \cite{troude2025illusions}, eigenvectors collapse onto a low-dimensional space as $\kappa\to\infty$,
making perturbations strongly aligned with the least stable mode.

Within a stochastic framework, this amplification manifests as a renormalization of effective temperature \cite{TroudeSornette2025},
\begin{equation} 
T_{\rm eff}\sim \left[1 + \left(\kappa/\kappa_c\right)^{2}\right]T~,
\label{thwygqb}
\end{equation} 
where $\kappa_c$ depends on the Jacobian spectrum,
so that Kramers escape rates across potential barriers gets renormalized as
\begin{equation}    \label{eq:Kramers-renorm}
     \Gamma \sim e^{-\Delta U_{\rm eff}/(k_BT_{\rm eff})},
\end{equation}
where $\Delta U_{\rm eff}$ is an effective barrier potential.
Non-normality dynamically lowers effective activation barriers, accelerating reaction transitions without altering the underlying thermodynamic potentials. As an illustrative example, consider a system with $k_{\mathrm{B}}T \sim 10^{-2}\,\Delta U_{\mathrm{eff}}$. In the normal case, the corresponding escape rate is astronomically small, $\Gamma \sim 4\times10^{-44}$, so that the system effectively remains trapped in its initial state. By contrast, for a moderate degree of non-normality, $\kappa \simeq 7$, the effective temperature is renormalized to $k_{\mathrm{B}}T_{\mathrm{eff}} \sim \Delta U_{\mathrm{eff}}/2$, yielding a substantial escape rate $\Gamma \sim 0.13$. This corresponds to an enormous acceleration of the reaction kinetics induced purely by non-normal dynamics.

\subsection*{Linking entropy production to kinetic acceleration}

In Methods, we show that the effective temperature renormalization (\ref{thwygqb})
is deeply related to the increase in entropy production (\ref{hh3umju3}) due to non-normality 
and can be reformulated as
\begin{equation}
T_{\rm eff} \sim (1+c\,\Phi)\, T ,
\end{equation}
where $c$ is a constant.
Pulling this back into an effective-landscape representation yields a modified free energy
\begin{equation}
F_{\rm eff} \;=\; U_{\rm eff} - T_{\rm eff}~ S~,
\end{equation}
where $U_{\rm eff}$ is the effective potential governing the slow degrees of freedom.
This construction embeds the NESS into an effective thermodynamic landscape whose curvature and barrier heights are dynamically reshaped by non-normal entropy production.
In this formulation, the acceleration of chemical reactions in open asymmetric networks arises from a renormalization of the energy landscape itself:
non-normal amplification increases $\Phi$, which raises $T_{\rm eff}$,
which in turn lowers effective barriers and facilitates fast transitions.

\subsection*{Non-normality as a physical mechanism for metabolic acceleration}

Together, these results establish the central claim:
\begin{itemize}[itemsep=0pt, parsep=0pt, topsep=0pt]
\item Asymmetric and directional structures in biochemical networks (non-reciprocal couplings, hierarchical feedbacks, and driven transport) distort the geometry of dynamical modes and thereby generate non-normality.
\item Non-normality increases eigenvector overlap and entropy production rate $\Phi$.
\item The same overlaps control transient amplification.
\item Amplification lowers effective kinetic barriers and accelerates chemical reaction rates.
\end{itemize}
Therefore, open, asymmetric chemical networks naturally exhibit fast, noise-triggered transitions, or ``pseudo-bifurcations'', long before true eigenvalue instabilities arise. 

This dynamical picture reframes a fundamental evolutionary truth: life is a fight between exponentials, where a minuscule advantage in long-term growth rate determines dominance. Non-normality provides a powerful weapon in this fight. By enabling the transient amplification of fluctuations and the acceleration of reaction kinetics, non-normal dynamics allow biological systems to effectively realize a higher growth rate, outpacing asymptotically stable competitors even when their eigenvalues are identical. This advantage in kinetic responsiveness underpins rapid adaptation and evolution. Crucially, this principle extends to fluctuating environments, where non-normality has been shown to increase the Lyapunov exponent \cite{troude2025kesten1,troude2025kesten2}---the long-term growth rate under multiplicative noise---suggesting a direct evolutionary pathway favoring the selection of non-normal network architectures.

\section*{Rationalizing Biological Acceleration across Scales via Non-Normal Dynamics}

To validate and illustrate our theory, we show how it rationalizes the function of core biological motifs. From molecular catalysis to cellular transport and regulatory networks, each example demonstrates a common causal chain: specific biological designs generate asymmetric couplings, leading to a non-normal operator with large $\kappa$, which in turn amplifies entropy production (scaling as $\kappa^2$ and accelerates dynamics. This consistent explanation across diverse systems underscores the framework's strength as a unifying physical principle.
 
As a first example, the classical Eyring-Arrhenius view treats enzymatic catalysis as a static lowering of the activation barrier.
Within the non-normal framework, however, part of this lowering emerges from dynamical transient amplification:
non-reciprocal couplings between the reaction coordinate and internal vibrational or electronic modes create a highly non-normal Jacobian for the enzyme-substrate complex.
Expression (\ref{eq:Kramers-renorm}) shows that the rate enhancement can be understood as an effective barrier reduction generated by transient growth, scaling approximately with the squared condition number $\kappa^2$.
This provides a unified explanation for extreme enzymatic accelerations ($10^{17}$-$10^{23}$-fold)~\cite{Knowles1991,Fersht1999,RadzickaWolfenden1995,Wolfenden2013}:
enzymes achieve extraordinary $\kappa$-values through asymmetric, evolutionarily sculpted couplings.
Thus, catalytic ``barrier lowering'' is not only thermodynamic but also a non-normal dynamical phenomenon that amplifies entropy production and reactive fluxes.

In turn, proteins exhibit hierarchical coupling between fast local motions and slow global modes.
Linearizing their dynamics as in \eqref{eq:lin_master},
reveals that the Jacobian $\A$
is generically non-normal due to directional and anisotropic intramolecular couplings.
Non-orthogonal eigenvectors mean that small perturbations in one region can undergo transient amplification into large-scale conformational shifts elsewhere.
In the non-normal framework, this hierarchical structure corresponds to large eigenvector overlaps, increasing the condition number 
$\kappa$ and thus entropy production by a factor proportional to $\kappa^2$.
This explains why allosteric transitions, long treated as special regulatory behaviors, arise naturally from the generic pseudospectral physics of proteins~\cite{Hyeon2011,Motlagh2014}.
Local energy inputs are dynamically amplified and propagated globally, revealing allostery as a non-normal mode-coupling effect that accelerates functional transitions and dissipative flux across the molecule.

Membrane-associated networks, such as electron transport chains and ATP synthases, are canonical directional systems.
Their reaction graphs contain irreversible couplings and gradient-driven cycles, violating reciprocity ($\A \neq \A^{T}$) and detailed balance.
As a result, the kinetic operator exhibits strong non-normality and supports persistent probability currents.
The steady-state entropy production, 
$ \Phi = \sum_{i,j} A_{ij} 
\ln \!\frac{A_{ij} p^{\ast}_{j}}{A_{ji} p^{\ast}_{i}} > 0$ is directly amplified by the antisymmetric component in the Lyapunov equation,
precisely the term enhanced by increased $\kappa$.
Thus, directional electrochemical systems naturally achieve high entropy production rates because their architecture maximizes non-normal eigenvector geometry.
This reframes classical bioenergetics~\cite{Nicholls2013,Qian2005}:
the efficiency and throughput of electron-proton coupling systems derive not only from thermodynamic gradients but also from dynamically amplified fluxes created by non-normal operators.
Electrochemical coupling therefore becomes a macroscopic realization of the same principle driving enzymatic catalysis and allostery:
increasing non-normality increases transient amplification, which increases entropy production, enabling life's accelerated energy transduction.

\section*{The Biotic--Abiotic Horse Race: Hydrogenotrophic Methanogenesis as a Test of Non-Normal Acceleration}

If life is a non-normal chemical accelerator, then under identical thermodynamic boundary
conditions it should perform the same redox chemistry as abiotic systems \emph{far faster} and
with \emph{tighter energetic coupling}. This prediction can be tested in the canonical
H$_2$--CO$_2$ system, where both biology and geochemistry reduce CO$_2$ with H$_2$ yet differ
enormously in kinetics.
In the Supplementary Information, we develop a simple model illustrating how
life can emerge as a dynamical transition in which ordinary abiotic chemistry is reorganized into a 
non-normal reaction architecture that amplifies fluxes, accelerates kinetics, and sustains nonequilibrium organization.

Hydrogenotrophic methanogens catalyze
\begin{equation}
    \mathrm{CO_2 + 4H_2 \rightarrow CH_4 + 2H_2O},
\end{equation}
using the MCR/F$_{430}$ complex embedded in an electron-transfer and proton-pumping
architecture that conserves the landscape approach as an ion motive force~\cite{Thauer2019_MCRreview,Muller2018_AnRevMicro,Dinh2024_Accounts}.
Abiotic analogues occur in serpentinizing systems: Fe(Ni)S membranes under pH gradients
produce formate at $\sim10^{-12}$~mol\,min$^{-1}$ per chip at room temperature~\cite{Hudson2020_PNAS},
and hydrothermal catalysts yield $\sim0.5$~mmol\,L$^{-1}$\,h$^{-1}$ of formate at 100~$^\circ$C~\cite{Preiner2020_NEE}.
By contrast, methanogens generate 9--28~mmol\,g$^{-1}$\,h$^{-1}$ of methane at 30--40~$^\circ$C,
equivalent to 4.5--28~mmol\,L$^{-1}$\,h$^{-1}$ for typical biomass densities~\cite{Goyal2015_MCF,Goyal2014_MSB}.
Thus, even at comparable temperatures, \emph{biotic rates exceed abiotic rates by 6--9 orders of magnitude}.

In the non-normal framework, this gap arises because biological reaction networks reorganize the same chemistry into architectures with much larger condition number $\kappa$, producing a dynamical amplification of reactive fluxes and entropy production that scales as $\kappa^{2}$.  
Abiotic systems, constrained by diffusion and lacking directional couplings or feedback,
exhibit $\kappa\approx1$: their fluxes remain limited by the bare thermodynamic drive.
Methanogens, by contrast, engineer strong asymmetries (vectorial electron transfer, membrane
potentials, gated ion flows) that yield large $\kappa$ and thus large
non-normal amplification, explaining their vastly superior turnover under the same redox
gradient. 

Coupling efficiency shows a parallel contrast.  
Cells recycle a substantial fraction of the energy barrier into an ion motive force and ATP synthesis,
while abiotic catalysts dissipate almost all energy as heat or neutral species.
In non-normal terms, biological systems use asymmetric couplings and boundary confinement to
channel part of the amplified dissipative flux into organized work.

The biotic--abiotic comparison therefore provides a sharp, falsifiable test of our principle:
if non-normal amplification is essential to life, then (1)~biotic rates should vastly exceed
abiotic rates under matched gradients; (2)~biological systems should conserve more of the
available free energy; and (3)~in mixed settings, active methanogens should drive fluxes and
isotopic signatures toward biotic values~\cite{Bradley2010_LostCity,Keir2010_GeophysResLett,Lang2012_LostCity}.
Current evidence supports all three predictions, suggesting that life is distinguished not
by its thermodynamic potential, but by its ability to construct \emph{high-$\kappa$,
non-normal architectures that accelerate chemical entropy flow}.


\section*{Conclusion and Outlook}


Across thermodynamics, statistical physics, biochemistry, and planetary science \cite{Kleidon2010_PhyLifeRev,SchwartzmanVolk1989Nature,SchwartzmanVolk1991,Porada2016NatCommun,Seitzinger2006EcolAppl,HuntSor25}, a single pattern emerges: life does not merely maintain disequilibrium, it \emph{accelerates} it.  From H$_2$+CO$_2$ redox chemistry to enzymatic catalysis, allostery, and membrane energetics, living systems consistently exploit \emph{non-normal} architectures in which asymmetric couplings and eigenvector overlap transiently amplify fluxes and enhance entropy production.  In this sense, life represents an ``accelerated phase of matter'': a self-organized network that converts free-energy gradients into the fastest chemical transformations.

We propose that the origin of life corresponds to a coupled kinetic-structure transition in which primitive reaction networks reorganize energy flow through non-normal dynamics.  Our synthesis suggests the following  physical pathway.  
(1) Primitive open reaction networks naturally operated in the non-normal regime.  
(2) Asymmetric couplings generated non-orthogonal modes and transient growth.  
(3) This amplification increased entropy production and throughput \emph{without eigenvalue instability}, enabling rapid chemistry while preserving stability.  
(4) Life emerged as the material embodiment of architectures that maximize this accelerated dissipation.

This perspective yields clear, testable predictions.  
(1) Entropy production and reaction rates should scale with the squared condition number $\kappa^{2}$ of the underlying kinetic operator.  
(2) Synthetic protocells or microreactors with tunable asymmetry should exhibit controllable transient amplification and coupling efficiency.  
(3) Evolutionary analyses should reveal selection for architectures that optimize eigenvector overlap to balance speed, robustness, and energetic return.

\begin{figure*}
    \centering
    \includegraphics[width=\textwidth]{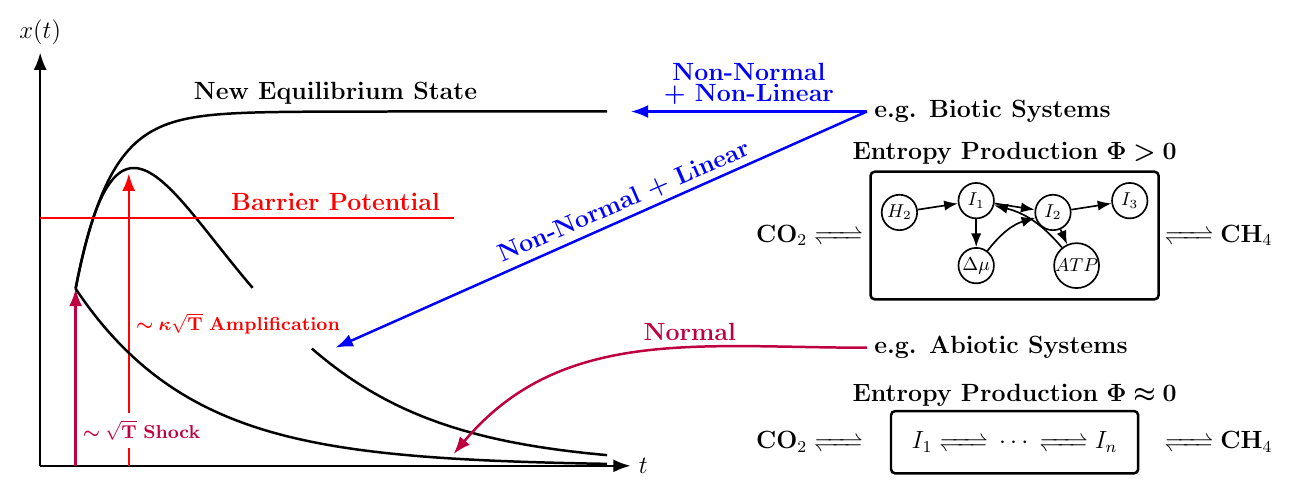}
    \caption{
        \textbf{Non-normal dynamics enable kinetic acceleration and barrier crossing in biotic chemical networks.}
        \textbf{Left | Dynamical response to a perturbation.}
        Shown is the time evolution of a coarse-grained reaction coordinate $x(t)$ following a small perturbation (``shock'') whose typical amplitude is set by thermal or stochastic fluctuations.
        In a normal (reciprocal) system, perturbations decay monotonically and exponentially back to the original equilibrium,
        otherwise you need a higher temperature to cross the barrier.
        By contrast, a non-normal but linearly stable system exhibits transient amplification:
        due to eigenvector non-orthogonality, initially small fluctuations of characteristic magnitude $\sim \sqrt{T}$ can grow transiently to amplitudes scaling as $\sim \kappa \sqrt{T}$,
        where $\kappa$ quantifies the degree of non-normality.
        When non-normal amplification acts in a purely linear regime,
        the system ultimately relaxes back to the original equilibrium.
        When combined with intrinsic non-linearities, however,
        the amplified fluctuation can cross the basin boundary of the initial state,
        overcome an effective barrier in state space, and relax into a new equilibrium.
        This transition occurs without any eigenvalue instability and is driven purely by the geometry of the dynamics and stochastic forcing. \\
        \textbf{Right | Chemical interpretation: biotic versus abiotic reaction architectures.}
        Schematic illustration of how the same net chemical transformation,
        here exemplified by the reduction of CO$_2$ to CH$_4$,
        is implemented in abiotic versus biotic systems.
        In abiotic chemistry (right), the reaction proceeds through a low-dimensional,
        approximately linear chain of intermediates with near-reciprocal couplings.
        The corresponding dynamics are close to normal, transient amplification is negligible,
        and the steady-state entropy production rate is small ($\Phi \approx 0$).
        In biotic systems (left), the same chemistry is embedded in a high-dimensional,
        driven reaction network involving multiple intermediates ($I_1, I_2, I_3,$ etc...),
        hydrogen supply (H$_2$), and energetic degrees of freedom such as membrane potential differences ($\Delta\mu$) and ATP turnover.
        Directional, asymmetric couplings and feedbacks render the effective kinetic operator strongly non-normal,
        leading to transient amplification of fluctuations, enhanced entropy production ($\Phi > 0$),
        and accelerated reaction kinetics. \\
        \textbf{Overall interpretation.}
        The figure illustrates the central idea that life acts as a \emph{non-normal chemical accelerator}:
        by reorganizing chemical reactions into asymmetric, energy-transducing networks,
        biological systems exploit non-normal dynamics to transiently amplify fluctuations,
        increase entropy production, and overcome kinetic barriers inaccessible to normal,
        abiotic chemistry, thereby achieving orders-of-magnitude faster reaction rates under comparable thermodynamic driving.
    }
    \label{fig:illustration}
\end{figure*}
 

\section*{Methods}

\subsection*{Model class and housekeeping entropy production}

We consider overdamped Langevin dynamics for concentrations or coarse-grained reaction
coordinates,
\begin{equation}
\dot{\x} = \f(\x) + \sqrt{2\delta}\,\xii(t),
\label{eq:Langevin_general}
\end{equation}
where $\x\in\mathbb{R}^n$, $\f(\x)$ is a generally non-conservative force field,
$\delta$ is the bare noise intensity (proportional to temperature),
and $\xii$ is standard Gaussian white noise.
The steady-state density $P_{\rm ss}(\x)$ solves the stationary Fokker--Planck equation with probability
current
\begin{equation}
\J_{\rm ss}(\x) = \f(\x)\,P_{\rm ss}(\x) - \delta\nabl P_{\rm ss}(\x).
\end{equation}

Steady-state thermodynamics (SST) decomposes the entropy balance into system entropy
$\dot S$ and an entropy flux $\Pi_h$ to the environment \cite{OonoPaniconi1998,HatanoSasa2001}.
For Langevin systems, the steady-state (housekeeping) entropy production rate is
\begin{equation}
\Phi \;=\; \frac{1}{\delta}\int \frac{\left\|\J_{\rm ss}\right\|^2}{P_{\rm ss}}\,dx,
\label{eq:Phi_general}
\end{equation}
which coincides with the housekeeping entropy flux $\Pi_{\rm h}$ in the NESS:
$\dot S=0$ and $\Phi=\Pi_{\rm h}$ \cite{Fyodorov2025}.
This object is what we call the entropy flux or entropy production rate throughout. 
It is strictly positive whenever the stationary current is non-vanishing, which occurs precisely when the force field contains a rotational (solenoidal) component.

\subsection*{Linear non-normal Langevin dynamics and eigenvector overlaps}

To expose how non-normality controls $\Phi$, we first linearize the dynamics around a
homogeneous steady state $\x^\ast$,
\begin{equation}
\delta\dot{\x} = \A\,\delta x + \sqrt{2\delta}\xii(t), 
\qquad \mw{\xii(t)\xii^\top(t')} = \mathbf{I}\delta(t-t'),
\label{eq:linear_Langevin}
\end{equation}
where $\A$ is the Jacobian at $\x^\ast$ of the force field $\f(\x)$.
Writing the spectral decomposition
$\A = \sum_i \lambda_i\rv_i\lv_i^\dag$ with biorthogonal left and right eigenvectors,
$\lv_i\cdot\rv_i=\delta_{ij}$, one obtains the compact expression \cite{Fyodorov2025}
\begin{equation}
\begin{split}
    &\qquad\qquad\Phi = \sum_{i,j} O_{ij}\,\Lambda_{ij}, \\
    &O_{ij} = \left(\rv_i\cdot\rv_j\right)\left(\lv_i\cdot\lv_j\right),
    \quad
    \Lambda_{ij} = \frac{\lambda_i(\lambda_i-\bar\lambda_j)}{\lambda_i+\bar\lambda_j}.
\end{split}
\label{eq:Phi_overlap}
\end{equation}
The quantities $O_{ij}$ are left-right eigenvector overlaps; they reduce to $O_{ij}=\delta_{ij}$
for normal matrices and become large when $\A$ is strongly non-normal.
The coefficients $\Lambda_{ij}$ depend only on the eigenvalues and are insensitive to eigenvector geometry.
Equation~\eqref{eq:Phi_overlap} makes explicit that non-normality enhances entropy production
through eigenvector non-orthogonality: at fixed eigenvalues, increasing overlaps $O_{ij}$ increases
$\Phi$.

\subsection*{Condition number, non-normal amplification, and linear scaling of $\Phi$}

Non-normal amplification is conveniently quantified by the condition number
$\kappa=\kappa(\PP)=\|\PP\|_2\|\PP^{-1}\|_2$ of the eigenvector matrix
$\PP=(\rv_1,\dots,\rv_n)$, or equivalently by pseudospectral quantities \cite{TrefethenEmbree2005}.

To obtain an explicit relation between $\Phi$ and $\kappa$, we use a two-dimensional linear
``$\beta$-form'' model \cite{troude2025unifying} that captures the essential non-normal coupling in the reduced subspace \cite{TroudeSornette2025}.
In appropriate units, the Jacobian can be written
\begin{equation}    \label{eq:beta}
\A =
\begin{pmatrix}
-1 & \beta\chi\kappa \\
\beta\chi/\kappa & -1
\end{pmatrix},
\end{equation}
where $\chi = 1$ for real eigenvalues, and $\chi=i$ for complex conjugate eigenvalues.
In both cases, $\beta$ is the distance to degeneracy, which occurs at $\beta=0$.
If $\chi=1$, then $\beta\in[0,1]$ also measures the distance to criticality:  the system approaches an instability as $\beta\to 1^-$.
If $\chi=i$, $\beta>0$ instead sets the oscillation frequency of the stable complex-conjugate eigenvalues.
The parameter $\kappa$ quantifies the degree of non-normality and is defined as the condition number of the eigenvector matrix $\PP$.

Using the form \eqref{eq:beta}, expression \eqref{eq:Phi_overlap} gives the following expression for the entropy production:
\begin{equation}    \label{eq:phi_K}
\Phi =
\begin{cases}
2(\beta K)^2, & \text{real eigenvalues,} \\
2\beta^2 (1+K^2), & \text{complex pair,}
\end{cases},
\quad\text{where } K = \frac{\kappa - \kappa^{-1}}{2}
\end{equation}
showing that in both cases $\Phi$ grows quadratically with the non-normality measure $K$ and
thus with $\kappa$ for large $\kappa$.
Complementarity, let us define the squared deviation of the system as
\begin{equation}
    v_\infty = \lim_{t\to\infty} \mw{\left\|\x(t)\right\|^2},
\end{equation}
where $\mw{\cdot}$ denotes expectations. 
Ref.~\cite{troude2025unifying} derived that
\begin{equation}
    v_\infty = 
    \begin{cases}
        \frac{\delta}{1-\beta^2}\left[1 + \left(\beta K\right)^2\right] &\text{if } \chi=1 \\
        \delta\left[1 + \frac{\left(\beta K\right)^2}{1 + \beta^2}\right] &\text{if } \chi=i
    \end{cases} .
\end{equation}
Using \eqref{eq:phi_K}, we obtain
\begin{equation}
    v_\infty = 
    \begin{cases}
        \frac{\delta_{\rm eff}}{1-\beta^2} &\text{if } \chi=1 \\
        \frac{\delta_{\rm eff}}{1+\beta^2} & \text{if } \chi=i
    \end{cases},
\end{equation}
where 
\begin{equation}
\delta_{\rm eff} = \delta \left(1 + \frac{1}{2}\Phi \right)
\end{equation}
defines an effective noise intensity that renormalizes, and amplifies, the bare noise amplitude $\delta$.
This expression serves as the linear prototype of the relation $\delta_{\rm eff}\sim \delta\,[1+c\,\Phi]$ used in the main text:
in linear systems, non-normal entropy production increases the effective noise felt along the reaction coordinate through this amplification mechanism.

\subsection*{Nonlinear entropy production and quasi-potential}

We now extend these results to nonlinear open reaction networks with non-normal Jacobian near
their NESS. Following Freidlin–Wentzell theory, we write the stationary density in WKB form
\begin{equation}
P_{\rm ss}(\x) \propto \exp\bigl[-Q(\x)/\delta\bigr],
\label{ghtnw}
\end{equation}
where $Q(\x)$ is the quasi-potential solving the stationary Hamilton-Jacobi equation at leading
order in $\delta$,
\begin{equation}
\f(\x)\cdot\nabl Q(\x) + \|\nabl Q(\x)\|^2 = 0 + O(\delta).
\label{eq:HJ}
\end{equation}
Inserting $P_{\rm ss}$ into \eqref{eq:Phi_general} and integrating by parts yields
\begin{equation}
\Phi 
= \frac{1}{\delta}\left(\mw{\|\f(\x)\|^2}
+ \mw{\|\nabl Q(\x)\|^2}\right)
+ 2\,\mw{\nabl\cdot \f(\x)},
\label{eq:Phi_nonlin_exact}
\end{equation}
where $\mw{\cdot}$ denotes expectation with respect to $P_{\rm ss}$.
Using the Hamilton-Jacobi equation \eqref{eq:HJ}, we can write
\begin{equation}
    \Phi 
    = \frac{1}{\delta}\mw{\f(\x)\cdot\left(\f(\x) - \nabl Q(\x)\right)}
    + 2\,\mw{\nabl\cdot \f(\x)},
\end{equation}

Using expression (\ref{ghtnw}), the stationary current can be written as 
\begin{equation}
J_{\rm ss}(x)=P_{\rm ss}(x)\,\big(f(x)+\nabla Q(x)\big),
\end{equation}
so that
\begin{equation}
\Phi 
= \frac{1}{\delta}\int P_{\rm ss}(x)\,\|f(x)+\nabla Q(x)\|^2\,dx.
\end{equation}
Decomposing the force
\begin{equation}
\f(\x) = -\nabl U(\x) + \f_{\rm sol}(\x),
\qquad \nabl\cdot \f_{\rm sol}=0,
\end{equation}
 yields
\begin{equation}
\Phi 
= \frac{1}{\delta}\int P_{\rm ss}(x)\,
\big\| f_{\rm sol}(x) + \nabla Q(x)-\nabla U(x)\big\|^2\,dx.
\end{equation}
In a purely variational system with detailed balance, where $f_{\rm sol}=0$ and $Q=U$, the stationary current vanishes identically and hence $\Phi=0$. 
Thus the housekeeping entropy production is entirely driven by the non-variational (rotational) component of the force field and by its mismatch with the gradient of the quasi-potential.

\subsection*{Fast-slow reduction and effective one-dimensional dynamics}

To make contact with reaction coordinates and the Kramers escape problem, we introduce a nonlinear
two-dimensional fast-slow model that generalizes the linear $\beta$-form and is aligned with the
non-normal direction $y$ and the reaction coordinate $x$:
\begin{equation}
\begin{aligned}
\dot x &= -\partial_x\phi_x(x) + \kappa\,\partial_y\psi_y(y) + \sqrt{2\delta}\,\xii_x(t),\\
\dot y &= -\partial_y\phi_y(y) + \kappa^{-1}\partial_x\psi_x(x) + \sqrt{2\delta}\,\xii_y(t).
\end{aligned}
\label{eq:fast_slow}
\end{equation}
Here $\phi_x,\phi_y$ are confining potentials,
$\psi_x,\psi_y$ generate non-potential shear-like couplings,
and $\kappa$ controls the non-normality of the Jacobian near the NESS \cite{TroudeSornette2025}. 
We assume a clear separation of time scales:
\begin{equation}
\partial_y\phi_y(y)\approx \alpha y,\qquad \alpha\gg 1,
\end{equation}
so that $y$ is a fast, strongly stable mode slaved to $x$.

Linearizing $\partial_y\psi_y(y)\approx \beta y$ and solving the $y$-equation as an Ornstein--Uhlenbeck
process yields, to leading order in $1/\alpha$,
\begin{equation}
y(t) \simeq \frac{1}{\alpha\kappa}\,\psi'_x\bigl(x(t)\bigr) + \sqrt{\frac{\delta}{\alpha}}\,\eta_y(t),
\end{equation}
with $\eta_y$ white noise.
Substituting into the $x$-equation one obtains an effective one-dimensional
Langevin dynamics
\begin{equation}
\dot x = -U'_{\rm eff}(x) + \sqrt{2\delta_{\rm eff}}\,\eta_x(t),
\label{eq:1d_effective}
\end{equation}
with
\begin{equation}
    \begin{split}
        & U_{\rm eff}(x) = \phi_x(x) - \frac{1}{\kappa_c}\,\psi'_x(x),
        \\
        &\delta_{\rm eff} = \delta\left[1 + \left(\frac{\kappa}{\kappa_c}\right)^2\right],
        \qquad
        \kappa_c = \frac{\alpha}{\beta}.
    \end{split}
\label{eq:Ueff_Deff}
\end{equation}
Non-normality thus has two coupled effects along the reaction coordinate: it tilts the effective
potential via the solenoidal coupling and it amplifies the effective noise by a factor
$1+(\kappa/\kappa_c)^2$. In this reduced description, the stationary density of $x$ is Gibbs-like,
\begin{equation}    \label{eq:px}
P(x)\propto \exp\left[-\frac{U_{\rm eff}(x)}{\delta_{\rm eff}}\right],
\end{equation}
such that the overall probability distribution can be approximated by
\begin{equation}
    \begin{split}
        &\qquad\qquad\qquad P(x,y) \approx P(x)P(y|x), \\
        &\text{where } P(y|x) = \sqrt{\frac{\alpha}{2\pi \delta}} e^{-\frac{\alpha}{2\delta}\left(y - \bar{y}(x)\right)^2}
        ,\; \bar{y}(x) = \frac{1}{\alpha\kappa} \psi_x '(x) .
    \end{split}
\end{equation}
Therefore, we can write the quasi-potential as
\begin{equation}
    Q(x,y) = \frac{U_{\rm eff}}{1+ \left(\frac{\kappa}{\kappa_c}\right)^2} + \frac{\alpha}{2}\left(y - \bar{y}(x)\right)^2 .
    \label{juj3enh2w}
\end{equation}

\subsection*{Nonlinear entropy production as a function of $\kappa$}

Using the fast--slow non-normal model, we evaluate the nonlinear housekeeping entropy production by
conditioning on $x$ and averaging over the fast variable $y$.
To do so, we introduce $y_0:= y_0 (x,y) = y - \bar{y}(x)$,
allowing us to write the force as
\begin{equation}
    f_x(x,y) = -U_{\rm eff}'(x) + \kappa\beta y_0
    ,\quad
    f_y(x,y) = -\alpha y_0 .
\end{equation}
Using the definition $\bar y(x) = (\alpha\kappa)^{-1}\psi'_x(x)$ and the shifted coordinate $y_0 = y - \bar y(x)$, 
differentiating (\ref{juj3enh2w}) with respect to $x$ and $y$ yields 
\begin{equation}
    \partial_x Q = \frac{U'_{\rm eff}}{1 + \left(\frac{\kappa}{\kappa_c}\right)^2} - \frac{1}{\kappa}\psi_x''(x) y_0 
    ,\quad
    \partial_y Q = \alpha y_0 .
\end{equation}
Recalling that the conditional mean and variance of $y_0$ are
\begin{equation}
    \mw{y_0}|_x = 0,\quad \mw{y_0 ^2}|_x = \frac{\delta}{\alpha},
\end{equation}
the entropy production rate conditional to $x$ can be written as
\begin{equation}
    \begin{split}
        \left.\Phi\right|_x &= 
        \frac{1}{\delta}\left.\mw{\f(\x)\cdot\left(\f(\x) - \nabl Q(\x)\right)}\right|_x - 2\left.\mw{\phi_x''(x) + \alpha}\right|_x \\
        &= \frac{\left(U_{\rm eff}'(x)\right)^2}{\delta}\left(1 + \frac{1}{1+\left(\frac{\kappa}{\kappa_c}\right)^2}\right) + \frac{\alpha}{\delta}\left(\frac{\kappa}{\kappa_c}\right)^2 \\
        &\qquad\qquad - \frac{1}{\kappa_c}\psi_x''(x) - 2 U_{\rm eff}''(x) .
    \end{split}
\end{equation}
The entropy production rate can then be obtained from $\Phi = \mw{\Phi|_x}$.
Noting that the marginal probability distribution of $x$ is given by \eqref{eq:px}, 
an integration by parts (assuming that the probability density vanishes at the boundaries) yields
\begin{equation}
\mathbb{E}\!\left[ U_{\rm eff}''(x) \right] 
= \frac{\mathbb{E}\!\left[\left(U_{\rm eff}'(x)\right)^{2}\right]}{\delta_{\rm eff}}.
\end{equation}
Using this identity, the entropy production rate can be written as
\begin{equation}
\Phi = \left(\frac{\kappa}{\kappa_c}\right)^2
\left[
\frac{\alpha}{\delta}
+ \mathbb{E}\!\left(\frac{U_{\rm eff}'(x)^2}{\delta_{\rm eff}}\right)
\right]
+ O\!\left(\frac{1}{\kappa_c}\right).
\label{eq:Phi_nonlin_intermediate}
\end{equation}
We have
\begin{equation}
\mathbb{E}\!\left[\frac{U_{\rm eff}'(x)^2}{\delta_{\rm eff}}\right]
= \delta_{\rm eff}\,I[P],
\end{equation}
where $I[P]$ is the Fisher information of the stationary density along the reaction coordinate.
In the Kramers regime, $U_{\rm eff}$ is locally quadratic and $\delta_{\rm eff} I[P]$ is of order unity,
reflecting the stiffness of the effective potential.
Putting these pieces together, the nonlinear entropy production takes the asymptotic form
\begin{equation}
\Phi \simeq \frac{1}{c}\left(\frac{\kappa}{\kappa_c}\right)^2,
\qquad
\frac{1}{c} = \frac{\alpha}{\delta} + \delta_{\rm eff}I[P],
\label{eq:Phi_kappa2}
\end{equation}
which is the nonlinear counterpart of the linear $\beta$-form scaling
and shows that non-normality amplifies entropy production by a factor $(\kappa/\kappa_c)^2$
without introducing higher powers.

Inverting Eq.~\eqref{eq:Phi_kappa2} and combining it with the result of \eqref{eq:Ueff_Deff} leads to the approximate relation
\begin{equation}
\delta_{\rm eff} \approx \delta\,[1 + c\,\Phi],
\label{eq:Deff_Phi}
\end{equation}
which links the renormalized noise scale, and thus the acceleration of chemical reactions \cite{TroudeSornette2025}, to the entropy production rate.

\subsection*{Effective Landscape Approach}

For non-normal systems, we have shown that the dynamics admits an effective potential $U_{\rm eff}$ and an effective noise scale $\delta_{\rm eff}\approx \delta(1+c\Phi)$, as given in \eqref{eq:Ueff_Deff}.
In stochastic thermodynamics, the noise scale is proportional to temperature, $\delta\sim T$, so this renormalization naturally defines an effective temperature
\begin{equation}
T_{\rm eff} = (1+c \Phi) T.
\end{equation}
Thus, non-normality reshapes the deterministic landscape through $U_{\rm eff}$ and, simultaneously, inflates the stochastic agitation through $T_{\rm eff}$.
These two ingredients allow us to recast the nonequilibrium steady state of a non-normal system into a familiar landscape framework, by defining an effective free energy
\begin{equation}
F_{\rm eff} = U_{\rm eff} -   T_{\rm eff}~S.
\end{equation}

Within this representation, the mechanisms of entropy production and kinetic acceleration acquire an intuitive thermodynamic interpretation.
When the effective potential dominates the entropic term $S T_{\rm eff}$, the system remains confined near the steady state.
But as non-normality increases, the amplified entropy production increases $T_{\rm eff}$, effectively lowering dynamical barriers and enabling the system to cross them even when the physical temperature $T$ is low.
In this sense, non-normality provides a dynamical route to barrier crossing, generating transitions, chemical reactions, or pseudo-phase transitions without requiring an actual increase in environmental temperature.

This effective-landscape perspective unifies our results: non-normality enhances entropy production, renormalizes the effective temperature, and accelerates reactions---placing non-normal nonequilibrium steady states within a natural extension of the classical equilibrium landscape picture.

\subsection*{Unified Abiotic-Biotic Model for the Emergence of Non-Normality}

The abiotic-biotic comparison in CO$_2$ + H$_2$ reduction provides an experimentally grounded setting in which to formalize the dynamical transition from normal,
diffusion-limited chemistry to the non-normal,
feedback-amplified organization characteristic of biological systems.
Existing abiotic models of CO$_2$ hydrogenation (metal-catalyzed or hydrothermal) \cite{Ueda2021,Jin2021,Quintana2023}
and biotic methanogenesis models \cite{Weedermann2013AnaerobicDigestion} share a common algebraic structure:
both are multi-step reaction networks
\begin{equation}
    \mathrm{CO_2} \;\rightleftharpoons\; I_1 \;\rightleftharpoons\; I_2 \;\rightleftharpoons\;\cdots\;\rightleftharpoons\; \mathrm{CH_4},
\end{equation}
where $I_k$ denote intermediate carbon states.
In the abiotic case, the network is shallow, nearly linear, and weakly coupled.
In the biotic case, the same backbone is augmented by electron-bifurcating modules \cite{Thauer2008MethanogenicArchaea},
membrane-coupled ion fluxes \cite{Weedermann2013AnaerobicDigestion},
and redox pools that feed back into earlier steps.
The key insight of our framework is that these two regimes can be written as
\emph{limits of the same effective dynamical operator}.

\medskip

\paragraph{Unified dynamical formulation}
Let $\boldsymbol{\rho}(t)=(\rho_1(t),\dots,\rho_N(t))$ collect all intermediate concentrations in the reaction network,
including (if present) energetic cofactors (ATP/ADP, F$_{420}$, ferredoxin),
ion motive force variables, or membrane potentials.
Both abiotic and biotic systems admit a representation
\(
    \dot{\boldsymbol{\rho}}=\f(\boldsymbol{\rho}),
\)
and near a stationary operating point $\x^\ast$ we linearize to obtain
\begin{equation}
    \dot{\x}= \A\,(\x-\x^\ast),
    \qquad\text{where}\qquad
    \A = \left.\mathbf{D}\f\right|_{\x^\ast}.
\end{equation}
The difference between abiotic and biotic chemistry is therefore encoded not in the eigenvalues of $\A$—which may be similar in magnitude—but in the \emph{pattern of couplings} inside $\A$.

\medskip

\paragraph{Abiotic limit (near-normal operator)}  
Abiotic CO$_2$ + H$_2$ reduction models (e.g.\ hydrothermal metal-catalyzed pathways)
consist of a nearly unidirectional chain with weak reversibility and negligible feedback into early steps \cite{Quintana2023}.  
This produces a Jacobian with weak anti-diagonal couplings, which 
has left/right eigenvectors that remain nearly orthogonal.  
In our framework this corresponds to a \emph{normal or weakly non-normal} operator i.e.
\(
\kappa^{\mathrm{abiotic}}/\kappa^{\mathrm{abioti}}_c\approx 1
\);
Such systems exhibit no significant transient amplification and therefore no geometric source of enhanced entropy production.
In the extreme case where the reaction can be collapsed into a single
effective step, then the Jacobian is essentially scalar and exactly normal.
Mathematically, this is obtained by:
\begin{itemize}
    \item projecting onto a lower-dimensional subspace
    (eliminate membrane variables, ATP, redox pools), and
    \item setting to zero the cross-coupling entries in $\A$.
\end{itemize}

\paragraph{Biotic limit (strongly non-normal operator)}
Biotic hydrogenotrophic methanogenesis augments the same reaction backbone with:
\begin{itemize}
    \item redox pools that couple exergonic and endergonic reactions,
    \item membrane-associated proton or sodium gradients,
    \item gating mechanisms (e.g. Mtr, Mcr/F$_{430}$),
    \item electron-bifurcation cycles introducing feedforward and feedback loops.
\end{itemize}
These additions introduce directed, asymmetric couplings and long-range feedback among intermediate states.  
The Jacobian therefore increases in dimension,
where the new dimensions capture the membrane or redox-induced interactions that have no abiotic analog.  
Such couplings can be expected to generically destroy normality and produce large eigenvector overlaps.  
In our framework, this corresponds to
\(
\kappa^{\mathrm{biotic}}/\kappa^{\mathrm{biotic}} _c\gg 1
\).
The emergence of large non-normality is the dynamical signature that the system has acquired a \emph{life-like architecture}:
it can amplify fluctuations, accelerate fluxes, and reorganize its internal chemical state purely through geometry rather than spectral instability.

\paragraph{Compatibility with calibration procedures}  
Because the unified model provides a single operator $\A$ for both abiotic and biotic regimes,
the same calibration pipeline applies.
Since both abiotic and biotic models map to the same reduced $\Gamma$,
one may directly compare their dynamical regimes and identify when biotic structure emerges.

\paragraph{Link to entropy production}  
Because entropy production depends on left-right eigenvector overlaps, the entropy-production machinery developed earlier applies without modification.  
In the abiotic limit, $\kappa\simeq 1$ and overlaps remain small, giving low $\Phi$; in the biotic limit, $\kappa\gg 1$, the overlaps inflate and $\Phi$ rises even under fixed thermodynamic driving.  
Thus the unified model provides the precise bridge between:

\begin{strip}
\begin{equation*}
\text{(i) reaction-network structure}
\quad\longrightarrow\quad
\text{(ii) non-normal operator geometry}
\quad\longrightarrow\quad
\text{(iii) entropy production}.
\end{equation*}
\end{strip}


\null\clearpage

\section*{Supplementary Information: Unified Abiotic-Biotic Model for the Emergence of Non-Normality}

Here, we develop a unified dynamical framework to describe the transition from abiotic to biotic chemistry through the emergence of non-normal reaction architectures. Using CO$_2$ + H$_2$ reduction as a common chemical backbone, we show that both abiotic hydrogenation pathways and biotic hydrogenotrophic methanogenesis can be represented by the same underlying dynamical operator, differing not in their spectra but in the pattern of couplings among reaction intermediates and energetic variables. In abiotic systems, weak feedback and near-reciprocal interactions yield nearly normal operators and diffusion-limited kinetics. In contrast, biotic systems augment the same reaction chain with redox pools, membrane-coupled ion fluxes, energy conservation modules, and biomass feedback, producing strongly non-normal Jacobians with large eigenvector overlaps. We demonstrate that this increase in non-normality enables transient amplification, accelerated chemical fluxes, and enhanced entropy production even under fixed thermodynamic driving. In this view, the emergence of life corresponds to a dynamical transition in which ordinary chemistry is reorganized into a non-normal chemical accelerator, providing a mechanistic link between biochemical architecture, kinetic enhancement, and sustained nonequilibrium organization.

    The abiotic-biotic comparison in CO$_2$ + H$_2$ reduction provides an experimentally grounded setting in which to formalize the dynamical transition from normal,
    diffusion-limited chemistry to the non-normal,
    feedback-amplified organization characteristic of biological systems.
    Existing abiotic models of CO$_2$ hydrogenation (metal-catalyzed or hydrothermal) \cite{Ueda2021,Jin2021,Quintana2023}
    and biotic methanogenesis models \cite{Weedermann2013AnaerobicDigestion} share a common algebraic structure:
    both are multi-step reaction networks
    \begin{equation}
        \mathrm{CO_2} \;\rightleftharpoons\; I_1 \;\rightleftharpoons\; I_2 \;\rightleftharpoons\;\cdots\;\rightleftharpoons\; \mathrm{CH_4},
    \end{equation}
    where $I_k$ denote intermediate carbon states.
    In the abiotic case, the network is shallow, nearly linear, and weakly coupled.
    In the biotic case, the same backbone is augmented by electron-bifurcating modules \cite{Thauer2008MethanogenicArchaea},
    membrane-coupled ion fluxes \cite{Weedermann2013AnaerobicDigestion},
    and redox pools that feed back into earlier steps.
    The key insight of our framework is that these two regimes can be written as
    \emph{limits of the same effective dynamical operator}.

    \medskip

    \paragraph{Unified dynamical formulation}
    Let $\x=(x_1,\dots,x_N)$ collect all intermediate concentrations in the reaction network,
    including (if present) energetic cofactors (ATP/ADP, F$_{420}$, ferredoxin),
    ion motive force variables, or membrane potentials.
    Both abiotic and biotic systems admit a representation
    \(
        \dot{\x}=\f(\x),
    \)
    and near a stationary operating point $\x^\ast$ we linearize to obtain
    \begin{equation}
        \dot{\x}= \A\,(\x-\x^\ast),
        \qquad\text{where}\qquad
        \A = \left.\mathbf{D}\f\right|_{\x^\ast}.
    \end{equation}
    The difference between abiotic and biotic chemistry is therefore encoded not in the eigenvalues of $\A$—which may be similar in magnitude—but in the \emph{pattern of couplings} inside $\A$.

    \medskip

    \paragraph{Abiotic limit (near-normal operator)}  
    Abiotic CO$_2$ + H$_2$ reduction models (e.g.\ hydrothermal metal-catalyzed pathways)
    consist of a nearly unidirectional chain with weak reversibility and negligible feedback into early steps \cite{Quintana2023}.  
    This produces a Jacobian with weak anti-diagonal couplings,
    and therefore has left/right eigenvectors that remain nearly orthogonal.  
    In our framework this corresponds to a \emph{normal or weakly non-normal} operator i.e.
    \(
    \kappa^{\mathrm{abiotic}}/\kappa^{\mathrm{abioti}}_c\approx 1
    \);
    Such systems exhibit no significant transient amplification and therefore no geometric source of enhanced entropy production.
    In the extreme case where the reaction is reduced to a single
    effective step, the Jacobian is essentially scalar and exactly normal.
    Mathematically, this is obtained by:
    \begin{itemize}
        \item projecting onto a lower-dimensional subspace
        (eliminate membrane variables, ATP, redox pools), and
        \item setting to zero the cross-coupling entries in $\A$.
    \end{itemize}

    \paragraph{Biotic limit (strongly non-normal operator)}
    Biotic hydrogenotrophic methanogenesis augments the same reaction backbone with:
    \begin{itemize}
        \item redox pools that couple exergonic and endergonic reactions,
        \item membrane-associated proton or sodium gradients,
        \item gating mechanisms (e.g. two essential gating and energy-coupling enzyme complexes in hydrogenotrophic methanogenesis: Mtr, Mcr/F$_{430}$),
        \item electron-bifurcation cycles introducing feedforward and feedback loops.
    \end{itemize}
    These additions introduce directed, asymmetric couplings and long-range feedback among intermediate states.  
    The Jacobian therefore increases in dimension,
    where the new dimensions capture the membrane or redox-induced interactions that have no abiotic analog.  
    We propose that such couplings generically destroy normality and produce large eigenvector overlaps.  
    In our framework this corresponds to
    \(
    \kappa^{\mathrm{biotic}}/\kappa^{\mathrm{biotic}} _c\gg 1
    \).
    The emergence of large non-normality is the dynamical signature that the system has acquired a \emph{life-like architecture}:
    it can amplify fluctuations, accelerate fluxes, and reorganize its internal chemical state purely through geometry rather than spectral instability.

\paragraph{Biotic hydrogenotrophic methanogenesis: high-dimensional, non-normal Jacobian}

Let us provide an explicit model.
For the biotic case, we consider hydrogenotrophic methanogenesis in archaea,
which executes the same redox chemistry CO$_2$+4H$_2 \to$ CH$_4$+2H$_2$O, but
via an enzyme-laden pathway that couples to membrane potential, redox pools,
ATP production and biomass growth. Detailed chemostat and energetic models
for this process have been developed in the context of anaerobic digestion,
gut methanogenesis and general thermodynamic--kinetic microbial models
\cite{Weedermann2013,FekihSalem2021,Ali2024,MunozTamayo2019,Jin2016}, and
metabolic--isotopic models of hydrogenotrophic methanogenesis have been
formulated by Cao et al.\ \cite{CaoBaoPeng2019} and related work
\cite{Gropp2021,Gropp2022,Liu2025}.

We augment the same chemical chain of intermediates $X_0,\dots,X_N$ as in the
abiotic case, but we add energetic, redox and biomass variables. A minimal
set is:
\begin{itemize}
\item $X_0,\dots,X_N$: same intermediates as abiotic; includes CO$_2$ and CH$_4$;
\item $E_1,\dots,E_K$: redox pools: e.g.\ F$_{420}$, ferredoxin, CoM/CoB, etc.;
\item $\Delta\mu_H$: membrane proton motive force or ion motive force;
\item $A$: ATP pool;
\item $Y$: biomass: methanogen population or active enzyme concentration.
\end{itemize}
Collecting these variables:
\[
u = (X_0,\dots,X_N)^\top \in \mathbb{R}^{N+1},
\qquad
q = (E_1,\dots,E_K,\Delta\mu_H,A,Y)^\top \in \mathbb{R}^{K+3},
\]
and defining perturbations around a stationary state,
\[
\delta u(t) = u(t) - u^*,\qquad
\delta q(t) = q(t) - q^*,\qquad
\delta x(t) =
\begin{pmatrix}
\delta u(t)\\
\delta q(t)
\end{pmatrix},
\]
the linearisation of the full nonlinear biotic model leads to the block system
\begin{equation}
\begin{pmatrix}
\dot{\delta u}\\
\dot{\delta q}
\end{pmatrix}
=
\underbrace{
\begin{pmatrix}
A & B \\
C & D
\end{pmatrix}
}_{J^{(\mathrm{bio})}}
\begin{pmatrix}
\delta u\\
\delta q
\end{pmatrix}
+
\eta^{(\mathrm{bio})}(t),
\label{eq:bio-block}
\end{equation}
where $J^{(\mathrm{bio})}$ is the biotic Jacobian, partitioned into:
\begin{itemize}
  \item $A \in \mathbb{R}^{(N+1)\times(N+1)}$, representing the chemical
    intermediate chain (which reduces to $J^{(\mathrm{abio})}$ when the
    couplings to energetic variables are removed);
  \item $D \in \mathbb{R}^{(K+3)\times(K+3)}$, representing internal feedbacks
    within the redox, membrane and ATP pools plus biomass dynamics;
  \item $B \in \mathbb{R}^{(N+1)\times(K+3)}$, representing feed-forward
    couplings from energetic variables back into the chemical chain
    (e.g.\ exergonic steps pumping protons, $\Delta\mu_H$ modulating reaction
    rates);
  \item $C \in \mathbb{R}^{(K+3)\times(N+1)}$, representing feedback of the
    chemical chain onto energetic and biomass variables.
\end{itemize}

A minimal yet explicit set of linearised equations, preserving the
chemical chain structure and the energetic couplings, is:
\begin{strip}
\begin{equation}
\boxed{
\begin{aligned}
\dot{\delta X}_0 &= -a_{00}\,\delta X_0 + a_{01}\,\delta X_1 
                   + b_{0\Delta}\,\delta\Delta\mu_H + \xi_0(t),\\
\dot{\delta X}_i &= a_{i,i-1}\,\delta X_{i-1} - a_{ii}\,\delta X_i 
                   + a_{i,i+1}\,\delta X_{i+1}
                   + \sum_{\ell=1}^K b_{i\ell}\,\delta E_\ell 
                   + b_{i\Delta}\,\delta\Delta\mu_H + \xi_i(t),\qquad i=1,\dots,N-1,\\
\dot{\delta X}_N &= a_{N,N-1}\,\delta X_{N-1} - a_{NN}\,\delta X_N
                   + b_{N\Delta}\,\delta\Delta\mu_H + \xi_N(t),\\[0.5em]
\dot{\delta E}_\ell &= c_{\ell 0}\,\delta X_0 + \sum_{j=1}^N c_{\ell j}\,\delta X_j 
                      - d_{\ell\ell}\,\delta E_\ell
                      + d_{\ell\Delta}\,\delta\Delta\mu_H + d_{\ell A}\,\delta A 
                      + \zeta_\ell(t),\qquad \ell=1,\dots,K,\\
\dot{\delta \Delta\mu}_H &= c_{\Delta 0}\,\delta X_0 + \sum_{j=1}^N c_{\Delta j}\,\delta X_j
                           + \sum_{\ell=1}^K c_{\Delta \ell}\,\delta E_\ell 
                           - d_{\Delta\Delta}\,\delta\Delta\mu_H
                           + d_{\Delta A}\,\delta A + \zeta_\Delta(t),\\
\dot{\delta A} &= c_{A0}\,\delta X_0 + \sum_{j=1}^N c_{Aj}\,\delta X_j
                 + \sum_{\ell=1}^K c_{A\ell}\,\delta E_\ell 
                 + c_{A\Delta}\,\delta\Delta\mu_H 
                 - d_{AA}\,\delta A + d_{AY}\,\delta Y + \zeta_A(t),\\
\dot{\delta Y} &= \alpha_0\,\delta X_0 + \sum_{j=1}^N \alpha_j\,\delta X_j 
                 + \alpha_A\,\delta A - \delta_Y\,\delta Y + \zeta_Y(t).
\end{aligned}
}
\label{eq:bio-explicit}
\end{equation}
\end{strip}

Here:
\begin{itemize}
  \item The $a_{ij}$ terms retain the abiotic chain structure:
    if $B=C=0$ and energetic / biomass variables are frozen, then $A$ reduces
    to $J^{(\mathrm{abio})}$ (up to small corrections).
  \item The $b_{i\ell},b_{i\Delta}$ couplings encode the fact that exergonic
    steps (e.g.\ methyl--CoM reduction and heterodisulfide reduction in
    hydrogenotrophic methanogens) pump protons and change the redox state
    of pools $E_\ell$, so that $\Delta\mu_H$ and $E_\ell$ feed forward into
    reaction rates of the chain.
  \item The $c_{\cdot j},c_{\cdot \ell}$ couplings encode the feedback of
    chain intermediates onto redox pools, membrane potential and ATP, as well
    as feedback of these energetic variables onto each other (e.g.\ ATP synthase
    consumes $\Delta\mu_H$; maintenance costs consume ATP, etc.).
  \item The biomass equation closes the loop: growth $\dot Y$ depends on
    fluxes through the chain and on ATP, while $Y$ in turn influences reaction
    rates via enzyme abundance (appearing in nonlinear form before linearisation).
\end{itemize}

In matrix form, the biotic Jacobian reads
\[
J^{(\mathrm{bio})}=
\begin{pmatrix}
A & B\\
C & D
\end{pmatrix},
\qquad
A \approx J^{(\mathrm{abio})},\quad B,C \neq 0.
\]

Even if all eigenvalues of $J^{(\mathrm{bio})}$ satisfy $\Re\lambda<0$, the
presence of sizable off-diagonal blocks $B$ and $C$ generically makes
$J^{(\mathrm{bio})}$ strongly non-normal. Writing the eigen-decomposition
\[
J^{(\mathrm{bio})} = P_{\mathrm{bio}} \Lambda P_{\mathrm{bio}}^{-1},
\]
one typically finds a large condition number
\[
\kappa_2(P_{\mathrm{bio}}) 
  = \frac{\sigma_{\max}(P_{\mathrm{bio}})}{\sigma_{\min}(P_{\mathrm{bio}})} \gg 1,
\]
where $\sigma_{\max}$ and $\sigma_{\min}$ are the largest and smallest singular
values of $P_{\mathrm{bio}}$. This is the hallmark of the ``non-normal chemical
accelerator'': strong transient amplification and persistent probability
currents in state space even when the system is linearly stable.

    \paragraph{Compatibility with our calibration procedure}  
    Because the unified model provides a single operator $\A$ for both abiotic and biotic regimes,
    the same calibration pipeline applies.
    Since both abiotic and biotic models map to the same reduced $\Gamma$,
    we may directly compare their dynamical regimes and identify when biotic structure emerges.

    \paragraph{Link to entropy production}  
    Because entropy production depends on left-right eigenvector overlaps, the entropy-production machinery developed earlier applies without modification.  
    In the abiotic limit $\kappa\simeq 1$ and overlaps remain small, giving low $\Phi$; in the biotic limit $\kappa\gg 1$, the overlaps inflate and $\Phi$ rises even under fixed thermodynamic driving.  
    Thus the unified model provides the precise bridge between:
    \begin{strip}
    \[
    \text{(i) reaction-network structure}
    \quad\longrightarrow\quad
    \text{(ii) non-normal operator geometry}
    \quad\longrightarrow\quad
    \text{(iii) entropy production}.
    \]
    \end{strip}

\null\clearpage

\bibliographystyle{naturemag}   
\bibliography{bibliography}  

\end{document}